# Large-Scale Sleep Condition Analysis Using Selfies from Social Media

Xuefeng Peng, Jiebo Luo, Catherine Glenn, Jingyao Zhan and Yuhan Liu

Department of Computer Science, University of Rochester
Rochester, New York 14627, USA
xpeng4@u.rochester.edu, jiebo.luo@rochester.edu
Department of Psychology, University of Rochester
Rochester, New York 14627, USA
catherine.glenn@rochester.edu

**Abstract.** Sleep condition is closely related to an individual's health. Poor sleep conditions such as sleep disorder and sleep deprivation affect one's daily performance, and may also cause many chronic diseases. Many efforts have been devoted to monitoring people's sleep conditions. However, traditional methodologies require sophisticated equipment and consume a significant amount of time. In this paper, we attempt to develop a novel way to predict individual's sleep condition via scrutinizing facial cues as doctors would. Rather than measuring the sleep condition directly, we measure the sleep-deprived fatigue which indirectly reflects the sleep condition. Our method can predict a sleep-deprived fatigue rate based on a selfie provided by a subject. This rate is used to indicate the sleep condition. To gain deeper insights of human sleep conditions, we collected around 100,000 faces from selfies posted on Twitter and Instagram, and identified their age, gender, and race using automatic algorithms. Next, we investigated the sleep condition distributions with respect to age, gender, and race. Our study suggests among the age groups, fatigue percentage of the 0-20 youth and adolescent group is the highest, implying that poor sleep condition is more prevalent in this age group. For gender, the fatigue percentage of females is higher than that of males, implying that more females are suffering from sleep issues than males. Among ethnic groups, the fatigue percentage in Caucasian is the highest followed by Asian and African American.

## 1 Introduction

In modern societies, with growing pressures from life and work, sleep condition has become increasingly a big concern to people. Individuals who encounter sleep disorders such as insomnia, sleep deprivation, sleep apnea, and so on may not only appear less healthy and attractive [6], but also suffer from poor physical and mental performances during the daytime [3]. To prevent and minimize the impairments caused by sleep disorders, many researchers endeavor in studying human sleep conditions. Traditionally, there are two popular types of methods. The first type is self-report based; one example of such type is the Pittsburgh Sleep Quality Index (PSQI) [8]. Another type is electronic device based; this type of methods measures individual sleep quality via digital PSG recording and scoring [11]. The former relies on individual's self-report, which may carry significant biases, while the latter requires complicated medical equipment and time. Such drawbacks prevent gaining deeper insights about human sleep conditions at a large scale.

Therefore, we are particularly interested in 1) finding a new way to predict sleep condition in a relatively easier and faster fashion, and 2) applying this new way to analyze the sleep conditions of the massive number of selfies on the social media.

With respect to our first goal, we understand that measuring the sleep condition directly could be difficult, thus we have decided to approach it indirectly by looking at sleep-deprived fatigue. It has been clinically identified that fatigue is one of the most common symptoms of poor sleep condition. A research paper has quantitatively associated sleep-deprived fatigue rate with human facial cues [20]. According to [20], sleep deprivation caused fatigue is heavily correlated with eight facial cues, including hanging eyelids, red eyes, dark circles under eyes, pale skin, droopy corner mouth, swollen eyes, glazed eyes, and wrinkles/lines around eyes. Figure 1, which is reprinted from [20], shows the correlations between the perceived fatigue rate and those facial cues. The rate from 0 to 100 indicates the degree of the facial cue from 'not at all' to 'very'.

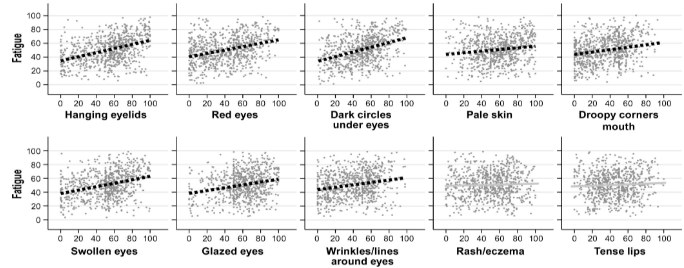

**Fig. 1.** Relationships between fatigue rates and the eight facial cues rates, reprinted from [20]. The first eight charts demonstrate that sleep-deprived fatigue correlates with the eight facial cues, whereas the last two show that sleep-deprived fatigue does not correlate with Rash/eczema and Tense lips.

Based on the correlation coefficients given in Figure 1, we constructed eight separate models (explained in section 2) to measure the eight facial cues for each individual. Next, with the measurement results, the overall fatigue rate can be quantified and treated as an indicator of this individual's sleep condition. Using these models along with our fatigue predicting criterion (explained in section 2.5), given a selfie, we can predict the perceived fatigue rate from the face within milliseconds, and with that fatigue rate we can preliminarily assess this individual's sleep condition.

We applied our prediction method on around 100,000 faces obtained from user timelines on Twitter and Instagram. For each face, we first utilized the *Face++* API[1], which is the state-of-the-art and reliable open-access face engine [4], to identify its demographic information and facial landmarks. We then located the interest areas that can indicate the facial cues through the landmarks. Those interest areas are sent to our models to produce an overall fatigue rate.

---
[1] http://www.faceplusplus.com

Our study found among the age groups (10-year per interval), fatigue percentage of 10-20 adolescent group is the highest, implying that poor sleep condition is more prevalent in this age group. This echoes the finding in [14] that sleep problems are common among adolescents. For gender, the fatigue percentage of females is higher than that of males, implying that more females are suffering from sleep issues than males. This result is consistent with the finding in [2]. Finally, among racial groups, the fatigue percentage in Caucasian group is the highest followed by Asian and African American.

Specifically, our main contributions include: 1) we develop a new method to predict sleep-deprived fatigue rate for a given face, 2) we apply our methodology at a large scale to study the population sleep condition using selfies from social media, and discover the correlations between sleep-deprived fatigue and demographic information.

## 2    Model Construction

Briefly, for each given face, we need to check eight facial cues that are highly correlated with sleep-deprived fatigue shown in Figure 1. To compute the fatigue rate according to the linear regression estimators given by Figure 1, the rate for each facial cue is also necessary. Therefore, we cooperated with the YR$^2$Lab in the Clinical and Social Psychology Department at the University, which provides the ratings of the eight facial cues for each training face. Next, we trained eight independent regression models where each of them will predict one corresponding facial cue rate. Subsequently, the combined linear regression estimator (Equation 1 in Section 2.3) is used to generate the overall fatigue rate.

### 2.1    Training Data Collection and Rating

Our most ideal training dataset would be a dataset that contains enough unique faces, some of which appear more fatigued than others, and some of which do not look fatigued at all. Moreover, for each unique face in the dataset, a few more facial images of that same face are also needed as references for the rating process (explained in section 2.2). Considering the expectations above, we have chosen the COLORFERET database[2] as our training dataset. The COLORFERET database is sponsored by NIST[3]. For each face, more than five high-quality facial imageries are provided. We picked 964 faces as our training dataset. Following Table 1 shows the distribution of age, gender, and race. Figure 2 shows the overall fatigue rate distribution.

---

[2] https://www.nist.gov/itl/iad/image-group/color-feret-database
[3] https://www.nist.gov

**Table. 1.** Age, gender, and race distribution of training set. Note that the average prediction confidences for gender and race are 95.2% and 90.5% respectively.

| Age | | | Gender | | Race | | |
|---|---|---|---|---|---|---|---|
| 0-20 | 20-40 | 40 - 60 | female | male | Asian | African American | Caucasian |
| 140 | 607 | 178 | 306 | 658 | 148 | 152 | 664 |

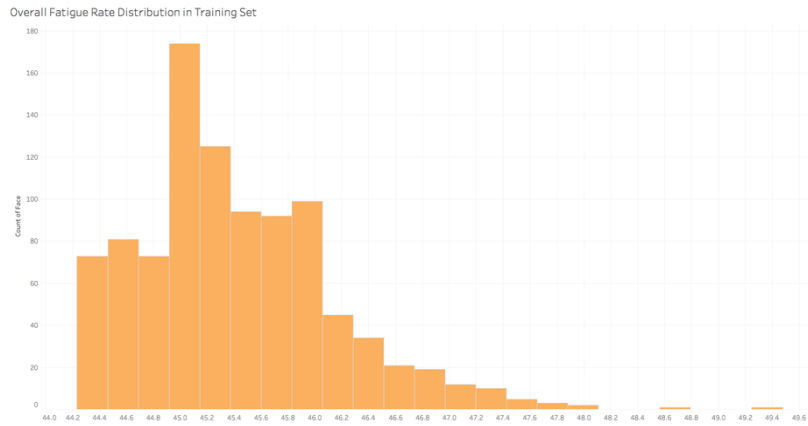

**Fig. 2.** Overall fatigue rate distribution for faces in training set.

For each training face, there are eight facial cues, to make the rating as objective as possible, we have invited three YR$^2$Lab[4] members with prior clinical research experiences to help us rate the training faces. All three raters have received an inter-judge agreement that describes the details about which area to scrutinize while rating a facial cue. Integers from 0 to 4 were used to indicate the rate of each facial cue with 0 means 'not at all' whereas 4 means 'very'. In addition, we applied three techniques to ensure the objectiveness of the ratings. First, each of the three raters was asked to rate eight facial cues for all 964 training faces, and the final rating for a facial cue of a face is calculated as the mean of the ratings given by the three raters. Second, sometimes the rater may be influenced by the previous facial image while rating the current one. Such influence could be significant if the display order is identical to all three raters. Therefore, we randomized the display order of the training faces for each rater to minimize such influence. Third, while displaying a face, the to-be-rated facial image is displayed, along with four or more images of the same face as references.

### 2.2 Feature Extraction for Facial Images

Six areas of interest are used to identify the eight facial cues of any given face. Left and right eye areas are examined to identify hanging eyelids, red eyes, swollen eyes, glazed eyes, and wrinkles/lines around eyes. The left and right eye bottom areas identify dark circle under eyes. The cheek area is for pale skin, and finally the mouth area is for droopy corner mouth.

---
[4] http://www.yr2lab.com/

For each training face, we first called the *Face++* API to obtain the facial landmarks of that face. We then cropped the areas of interest from the original facial image accordingly. The feature extraction algorithm we employed is Dense SIFT [1]. This algorithm will generate a feature descriptor for each cropped image. Note that for eye-related areas of interest, we concatenated the feature descriptors of the left and right eyes as a single feature descriptor, which we name eye feature descriptor; and we created the eye bottom feature descriptor in the same fashion.

### 2.3 Model Training and Prediction

We have eight facial cues to rate for any given face, and the rate of each facial cue depends on one corresponding feature descriptor. Therefore, we built eight separate models to predict the rate for each facial cue.

We fit our data by an ensemble of regression leaners. Prior to the training process, we standardized each feature descriptor. Next, we trained each model with the corresponding feature descriptors and the ground truth facial cue rates.

We performed Bayesian Optimization [19] to optimize the hyper-parameters of our models. The optimization aims to locate the combination of 1) ensemble-aggregation method, 2) number of ensemble learning cycles, 3) learning rate for shrinkage, and 4) minimal number of leaf node observations that produces the least 5-fold cross validated RMSE. The following Table 1 shows the best hyper-parameters for each model found within 30 iterations, along with the RMSE rates.

**Table 2.** Parameters and performance of models. P1, P2, P3, and P4 represents method, learnCycle, learnRate, and minLeafSize respectively. $X_i$ denotes the model for facial cue $x_i$.

| mdl | P1 | P2 | P3 | P4 | RMSE | mdl | P1 | P2 | P3 | P4 | RMSE |
|---|---|---|---|---|---|---|---|---|---|---|---|
| X1 | LSBoost | 499 | 0.011 | 122 | 1.825 | X5 | Bag | 340 | - | 27 | 1.747 |
| X2 | LSBoost | 50 | 0.090 | 146 | 1.651 | X6 | Bag | 195 | - | 1 | 2.056 |
| X3 | LSBoost | 86 | 0.065 | 67 | 1.983 | X7 | Bag | 118 | - | 59 | 1.895 |
| X4 | Bag | 494 | - | 481 | 1.745 | X8 | LSBoost | 499 | 0.017 | 42 | 2.033 |

Those eight models together constitute of our final model, which we call composite model, and it produced a 1.16% 5-fold cross validated SMAPE[5] on overall fatigue rates.

We located the coordinates of each linear regression estimator (black dot line) in the charts given by Figure 1 to derive its mathematical expression. Since the ranges for the fatigue rate and each facial cue rate are equivalent, we therefore can derive a combined linear regression estimator to compute the overall fatigue rate by averaging those eight models as

$$y = 0.037x_1 + 0.03x_2 + 0.041x_3 + 0.014x_4 + 0.022x_5 + 0.033x_6 + 0.027x_7 + 0.024x_8 + 44.41 \qquad (1)$$

---

[5] $\text{SMAPE} = \frac{2}{n}\sum_{t=1}^{n}\frac{|F_t - A_t|}{|F_t| + |A_t|}$

Eight variables $x_1, x_2, x_3, x_4, x_5, x_6, x_7,$ and $x_8$ represent the rate of hanging eyelid, red eye, dark circle, pale skin, droopy corner mouth, swollen eye, glazed eye, and wrinkles, respectively. The following is the procedure outline for predicting the overall fatigue rate for any given face: 1) obtain the facial landmarks, 2) crop the interest areas and run corresponding models to obtain the rate of each facial cue, 3) compute the overall fatigue rate through Equation 1.

## 3    Selfie Collection, Processing and Prediction

There are several keywords that people frequently tag when they post selfies on social media. They are "#selfie", "#me", "#happy", "#fun", "#smile", "#nomakeup", "#friends", and "#family". We used those tags to search the photo posts on Twitter and Instagram. For each post, we acquired a uid (user Id), and such uid can be utilized to backtrack the uid owner's timeline posts, which are with keywords.

It is possible that photos in a user's timeline contain not only his/her own faces, but also the faces of his/her friends, family members, and even strangers. Therefore, we grouped the faces into distinct face sets. Subsequently, we identified age, gender and race, located the facial landmarks of every single face in the face sets, then the areas of interest were extracted and the feature descriptors were generated. Finally, the overall fatigue rate of each face is predicted. The algorithms we have used for face detecting, facial landmarks locating, and face grouping are provided by *Face++* API.

## 4    Main Result

The demographic distributions of the social media selfies we have collected are summarized in the following Table 3.

**Table 3.** Demographic distribution. The average prediction confidence for gender and race are with 94.6% and 86.2%, respectively.

| Age (10-year per interval) | | | | | | | Gender | | Race | | |
|---|---|---|---|---|---|---|---|---|---|---|---|
| 0-1 | 1-2 | 2-3 | 3-4 | 4-5 | 5-6 | 6-7 | F | M | A | AA | C |
| 7624 | 23666 | 34251 | 23814 | 5678 | 2587 | 1096 | 58129 | 39033 | 17394 | 6168 | 73600 |

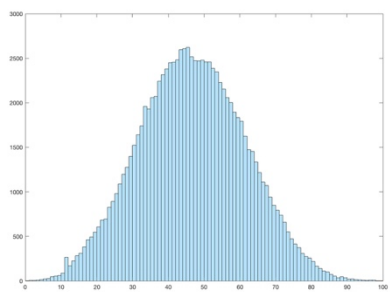

(a)

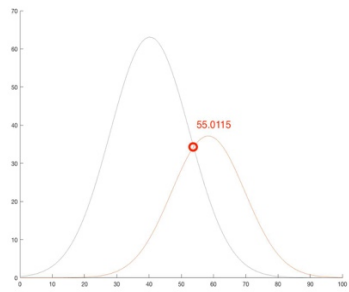

(b)

**Fig. 3.** The overall Fatigue rate distribution in (a), and the fitted Gaussian distributions in (b).

To see the overall fatigue rate differences more clearly, we normalized the overall fatigue rates into [0,100]. To pick the threshold that classifies faces into the fatigue set and non-fatigue set, we fitted our data into two Gaussian distributions. The right chart in Figure 3 shows the two distributions. We have chosen their intersection 55.0115 as our classification threshold (circled in red in Figure 4).

Using this threshold, we found that among all the unique faces we have collected, 29.09% faces are classified into the fatigue class, and the remaining 70.91% faces are classified into the non-fatigue class.

### 4.1 Age, Gender, and Race

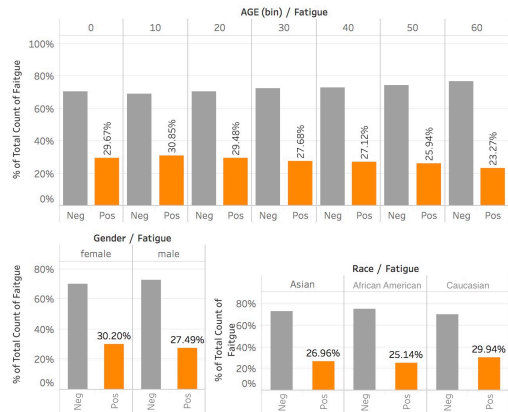

**Fig. 4.** Age, gender and race proportions on fatigue.

**Table. 4.** The details of the statistical significance test on fatigue proportion difference among age groups. According to [12], the difference is significant if the critical value exceeds the critical range. Note we only report significant differences. 0 stands for 0 – 10, 1 stands for 10 – 20, so forth.

| age   | critical val. | critical range | 95% CI (%)    | age   | critical val. | critical range | 95% CI (%)     |
|-------|---------------|----------------|---------------|-------|---------------|----------------|----------------|
| 0 - 3 | 0.021         | 0.020          | 1.05 – 3.24   | 1 - 4 | 0.037         | 0.023          | 2.49 – 5.00    |
| 0 - 4 | 0.027         | 0.026          | 1.22 – 4.17   | 1 - 5 | 0.055         | 0.031          | 3.82 – 7.26    |
| 0 - 5 | 0.045         | 0.034          | 2.61 – 6.37   | 1 - 6 | 0.076         | 0.046          | 4.97 – 10.19   |
| 0 - 6 | 0.065         | 0.048          | 3.81 – 9.25   | 2 - 5 | 0.036         | 0.030          | 1.94 – 5.31    |
| 1 - 2 | 0.019         | 0.013          | 1.19 – 2.63   | 2 - 6 | 0.057         | 0.046          | 3.08 – 8.26    |
| 1 - 3 | 0.032         | 0.014          | 2.40 – 3.99   |       |               |                |                |

In terms of age, more people in the youth and middle-age groups tend to appear fatigued and thus suffer from worse sleep conditions, one possible causation could be higher rates of anxiety and stress among these groups. The fatigue percentage of 0 - 20 age group is the highest, as the fatigue proportion difference between 0-10 and 10-20 groups is statistically insignificant. We employed Marascuilo procedure [12] to conduct the multiple proportions test, and the test result suggests that the proportion differences between 0-20 age group and other groups are statistically significant. An explanation is that many adolescents in this age group appear fatigue mainly due to their school workload or their frequent stay-up-late behaviors. A worldwide research on adolescent sleep [14] reveals that sleep problems are common among adolescents and many countries have reported high incidences of sleep disturbances in these age groups; and this outcome further confirms this notion using a large-scale data set that is obtained unobtrusively. We reported the critical value, critical range, and 95% confidence interval in Figure 5 (b).

With respect to gender, the proportion test indicates that the fatigue percentage of female is 2.1% to 3.3% higher than that of male with 95% confidence. This result echoes one of the findings in the 2005 Sleep in American poll summary [2] that females are more likely than males to have difficulty falling and staying asleep and thus more likely to experience more daytime fatigue. Among racial groups, our result suggests that the fatigue percentage in Caucasian group is the highest followed by Asian and African American. The Marascuilo procedure suggests that the proportion differences among those racial groups are statistically significant with 95% confidence interval of differences between Asian and African American, Asian and Caucasian, African American and Caucasian are 0.5% to 3.0%, -3.1% to -1.6%, and -5% to -3.1%, respectively.

## 5    Related Work

Our work builds upon previous research on sleep condition, fatigue studies, and computer vision. In sleep condition and fatigue studies, several studies have pointed out that fatigue is one of the most common symptoms of sleep deprivation. Furthermore, researchers also have found that the sleep-deprived fatigue rate is correlated with eight facial cues [20], and our study is based on the assumption that sleep-deprived fatigue is mainly reflected by those eight facial cues, and the fatigue rate can imply sleep condition. In terms of computer vision research, our work is related to face detection [16], gender, race and age identification [5], facial landmarks location [15], face grouping [21], and using visual social media to monitor mental health [10].

## 6    LIMITATIONS AND FUTURE WORK

Our work is built on the assumptions that the sleep-deprived fatigue can indicate one's sleep condition, and that sleep-deprived fatigue is associated with the eight facial cues. However, it is possible that the fatigue appearance is caused by some other chronic diseases [13] or a day of extremely heavy labor. Note that the use of social media selfies as the sensory data partially mitigates the above factors because people in those conditions are unlikely to post selfies. In addition, some people may naturally appear more fatigue than others thus using fatigue to infer their sleep conditions could be biased.

Therefore, we plan to develop more robust techniques to establish the face appearance *baseline* for a given individual before inferring his/her sleep condition. In addition, the overall fatigue rates of individuals who wear make-up can be underestimated. That is why we included "#nomakeup" as one of our keywords while retrieving selfies. Lastly, due to the restriction in accessing the original data in [20], Equation (1) is solely derived by averaging those eight linear estimators presented in Figure 1, which may lead to inaccurate measurement of the impact each facial cue constitutes for the overall fatigue rate. Nonetheless, we believe the trends and distributions in our study will remain consistent and valid, especially given that a large number of facial images are analyzed.

Our current research primarily focuses on studying the sleep condition distribution regarding to the demographic information. In the future, we may consider ways to identify the occupations of the face owners to investigate the distribution among different occupations. A recent study [9] has proposed a way to identify if a Twitter user is a college student or not. This method may help us unveil more interesting fatigue patterns of the student population. Moreover, we are also interested in examining the fatigue distribution over different geographical areas.

# 7 CONCLUSIONS

We have developed a new method to gauge a face's overall fatigue rate and use this rate to predict the face owner's sleep condition. This leads to a data-driven methodology to include a massive number of faces on social media to obtain the fatigue distributions with respect to age, gender, and race. Our main findings are largely consistent with those reported by using conventional small-scale empirical studies. This is extremely encouraging as it validates the effectiveness of the data-driven approach to study public health at large scales. Some of our findings are beyond those reported in the literature, e.g., those related to the interplays of age, gender and race, pointing to the potential to discover factors that the empirical studies have overlooked. Moreover, analyzing social media posts and pictures offers the potential to provide a method for mass screening for individuals at risk for a range of poor health conditions. Social media has been used to pick up signals when individuals use language that could be related to risky behaviors [17]. Tracking pictures could work in the same way to detect risk and allow for early intervention. Our work is in the same vein as [17, 18] and such social media-driven methods are expected to find more successes in computational psychology.

**Acknowledgement.** We thank the support of New York State through the Goergen Institute for Data Science, and our corporate research sponsors Xerox and VisualDX.